\documentclass[11pt]{article}
\usepackage{epsfig}
\begin{document}  
\thispagestyle{empty}
\begin{flushright}
hep-lat/0402008
\end{flushright} 
\vskip10mm
\begin{center}
{\LARGE
Topological lumps and Dirac zero modes in\\
\vskip2mm
SU(3) lattice gauge theory on the torus}
\vskip11mm
{\bf Christof Gattringer and Rainer Pullirsch}
\vskip3mm
Institut f\"ur Theoretische Physik, Universit\"at
Regensburg \\
D-93040 Regensburg, Germany 
\vskip35mm
\begin{abstract}  
We compute eigenmodes of the lattice Dirac operator for
quenched SU(3) gauge configurations on the 4-torus
with topological charge $\pm 1$. We find a strong dependence of 
the zero modes on the boundary conditions which we impose for the 
Dirac operator. The lumps seen by the eigenmodes often change their position
when changing the boundary conditions, while the local
chirality of the lumps remains the same. Our results show that the
zero mode of a charge $\pm 1$ configuration can couple to more than 
one object. We address the 
question whether these objects could be fractionally charged lumps. 
\end{abstract}
\end{center}
\newpage
\setcounter{page}{1}
\noindent
{\bf Introduction}
\vskip2mm
\noindent
Understanding the nature of the QCD vacuum still is an open and challenging 
problem. In particular, a possible connection between confinement and chiral 
symmetry breaking seems plausible. A hint that such a connection might 
exist can be seen in the fact that at the QCD phase transition deconfinement 
and chiral symmetry restoration take place at the same temperature. 

A step in the direction of getting a grip on such a mechanism was
the finding of so-called Kraan - van Baal (KvB) solutions \cite{kvb}. 
KvB solutions
are self dual solutions of the classical Yang-Mills equations 
on a euclidean cylinder (S$^1 \times \mbox{I\hspace{-0.8mm}R}^3$).
Thus they correspond to SU(3) gauge theory\footnote{Actually the KvB 
solutions are known for all SU(N), but here we only address the case 
of SU(3).} with temperature. A remarkable
property of KvB solutions is that they consist of 3 sub-lumps and these
three constituents together build up an object of topological charge 1. 
Furthermore these lumps were found to be BPS monopoles, a property 
that might lead to the possible connection to the confinement mechanism. 
Recently also KvB solutions with higher topological charge were found
\cite{highercharge},
giving rise to $3 |Q|$ monopoles for the solutions with topological 
charge $|Q|$.

Strong evidence that KvB solutions play a role also in the quantized theory
was found by several studies on the lattice. 
For SU(2) analyses based on cooling with twisted boundary 
conditions \cite{kvblattice1}
and with regular boundary conditions \cite{kvblattice2}
provided configurations
that show the features of KvB solutions. The existence of 
constituents was established and the characteristic behavior of the 
Polyakov loop was seen. The zero modes of the Dirac operator showed 
the expected change of location as a function of the boundary condition.
For SU(3) evidence
for KvB solutions comes from a study of eigenvectors of the Dirac operator
\cite{kvblattice3,kvblattice4} in thermalized (not cooled) configurations. 
In \cite{kvbzeromodes} the zero mode of the Dirac operator in the background   
of a KvB soltion was computed and it was found that it strongly depends on 
the fermionic boundary condition used. The zero mode is 
localized on only one of the lumps but the particular constituent 
it chooses is determined by the boundary condition. This behavior of the 
zero mode was found for thermalized configurations in 
\cite{kvblattice3,kvblattice4} and good
agreement with the properties of KvB zero modes was established. 
It is furthermore known that the gauge lumps are predominantly (anti-)
self dual \cite{selfdual} and a first
step towards matching the results from cooling with fermionic methods
is documented in \cite{kvblattice5}.

The analysis of Dirac eigenmodes in thermalized configurations makes use
of the fact that the low lying modes couple only to the long range 
structures of the gauge field and thus are an efficient filter for 
removing the 
UV fluctuations. Allowing for more general fermionic boundary conditions
(in particular an arbitrary complex phase) makes this method even more 
powerful and led to establishing the KvB behavior of zero modes in finite
temperature gauge configurations. 

It is natural to apply the generalized boundary condition technique 
also to gauge ensembles at zero temperature. A first step in this direction
was made in \cite{kvblattice4} where for a few configurations with topological 
charge $\pm 1$ it was shown that again the zero mode can be localized on 
different lumps when changing the boundary conditions. This hints at 
the existence of constituents also for charge $\pm 1$ configurations on 
the 4-torus T$^4$. For this manifold no analytic solution is known
(see \cite{t4analytic} for a discussions of the classical Yang-Mills
fields on the 4-torus) and a 
lattice study has a more exploratory aspect in this case. However, for
the case of two compact directions 
(T$^2 \times \mbox{I\hspace{-0.8mm}R}^2$) the recently found
analytic solution \cite{fopa}
again shows constituents and is already one step closer 
to the 4-torus analyzed here. 

In this article we present a large scale study of the Dirac eigenmodes 
for thermalized SU(3) configurations with topological charge $\pm 1$ 
on the 4-torus and analyze the 
dependence of the eigenmodes on the boundary conditions. We compare 
ensembles at two different volumes to analyze finite size effects. 
We show that in both volumes about 40\% of the configurations have 
zero modes that change their position when switching from periodic to
anti-periodic boundary conditions. This finding indicates that also 
on the torus a large 
portion of charge $\pm 1$ configurations gives rise to zero modes that 
couple to more than one object. 

An important question is whether these topological lumps have 
fractional charge and are truly constituents of a charge 1 object. 
The fact that the different lumps show up in the one and only zero mode 
we see supports this interpretation. Also the KvB solutions 
and a finite volume argument we discuss give hints in this direction. 
\vskip5mm
\noindent
{\bf Description of observables}
\vskip2mm
\noindent
For our calculations we use two ensembles of quenched gauge configurations
generated with the L\"uscher-Weisz gauge action \cite{luweact}
on lattices with sizes $12^4$ 
and $16^4$. For both lattices we use a coupling of $\beta = 8.45$ which 
gives rise to a lattice spacing of $a = 0.094(1)$ as determined in 
\cite{scale} from the Sommer parameter. For the gauge fields the 
boundary conditions are periodic.

We compute eigenvectors $\psi$ of the chirally improved lattice Dirac 
operator \cite{chirimp} which has good chiral properties and
was shown \cite{instantontest} to reproduce the zero mode
of a discretized instanton down to radii relatively
small in lattice units. In our calculation we allow for a general phase 
at the temporal boundary of the lattice Dirac operator.
The gauge fields have periodic boundary conditions and they
do not single out a time direction so we choose the 4-direction 
to be the time-direction where we impose the general boundary condition
for the Dirac operator and its eigenmodes
\begin{equation}
\psi(\vec{x},L_4+1) \; \;  \; = \; \; \; e^{i2\pi \, \zeta} \; \psi(\vec{x},1)
\; .
\end{equation}
The boundary phase parameter $\zeta$ can assume values in the interval [0,1]
with the two endpoints 0 and 1 giving periodic boundary conditions. 
Anti-periodic temporal
boundary conditions correspond to $\zeta = 0.5$.  

For all gauge configurations generated we compute 50 (30 for $16^4$)
eigenvalues and eigenvectors with periodic boundary conditions 
($\zeta = 0$) using the Arnoldi method \cite{arnoldi}. 
From the complete ensemble we then select those configurations which 
have exactly one zero-mode, i.e.\ have topological charge $Q = \pm 1$ 
according to the index theorem. For this subset of configurations we 
then solve the eigenvalue problem a second time now using anti-periodic
boundary conditions ($\zeta = 0.5$). The restriction to configurations with 
topological charge $Q = \pm 1$ avoids possible problems with mixing of 
the zero modes and allows for a clean interpretation of the findings. 
Our statistics is 291 configurations with charge $Q = \pm 1$
for $12^4$ and 199 configurations for $16^4$.
For a small subsample we also 
analyzed the eigenvalue problems with increasing $\zeta$ in small steps
$\zeta = 0.0, 0.1, 0.2 \; ... \; 0.9$. 

From the eigenmodes $\psi$ we can construct the scalar density 
$\rho$  and the pseudoscalar density $\rho_5$ as 
\begin{eqnarray}
\rho(x) & = & \sum_{\alpha,c} \;
\psi(x)_{\alpha,c}^\star \, \psi(x)_{\alpha,c} \; \; ,
\label{scalardens} 
\\
\rho_5(x) & = & \sum_{\alpha,\alpha^\prime,c} \;
\psi(x)_{\alpha,c}^\star \, (\gamma_5)_{\alpha,\alpha^\prime} 
\psi(x)_{\alpha^\prime,c} \; \; .
\label{pseudodens}
\end{eqnarray}
The star denotes complex conjugation. Both densities are gauge invariant
observables. The scalar density
$\rho$ is always non-negative, while the pseudoscalar density 
can have both signs, i.e.\ it is sensitive to the chirality of the lumps.
Note that for the scalar density we have 
$\sum_x \rho(x) = 1$, since the eigenvectors are normalized to 1.

As we have discussed above, we find localized structures in the 
eigenmodes, i.e.\ localized lumps in $\rho$ and $\rho_5$. A first 
observable is the position $x^{max}$ of the peak of the lump seen in 
the densities. As announced the eigenmode can be located at different
positions when comparing different boundary conditions. We will be
particularly interested in the distance between the maxima seen by
the periodic and the anti-periodic zero mode which is given by
\begin{equation}
d \; \; \; = \; \; \; \parallel \; x^{max}_{periodic} \; - \; 
x^{max}_{anti \,p.} \; \parallel_{torus} \; .
\label{distance}
\end{equation}
Here $\parallel .. \parallel$ denotes the 4-dimensional euclidean distance. 
The subscript $torus$ indicates that we always take the shortest distance 
possible on the 4-torus.
  We also find that the eigenmodes come in different sizes ranging from
tall and narrow (localized) to wide spread (delocalized). 
A convenient measure for the localization of an eigenmode is its 
inverse participation ratio $I$. It is defined as 
\begin{equation}
I \; \; = \; \; V \; \sum_x \rho(x)^2 \; .
\label{iprdef}
\end{equation}
Here $V$ denotes the number of lattice points.
It is easy to verify that for a maximally localized lump
($\rho(x) = \delta(x,x_0)$) one finds $I = V$, while for a maximally 
spread out lump ($\rho(x) = 1/V$) one obtains $I=1$. Thus a large value of 
$I$ corresponds to a localized, i.e.\ 
tall and narrow peak, while small values of $I$ come from 
widely spread, delocalized modes.
\vskip5mm
\noindent
{\bf Properties of zero-modes with general boundary conditions}
\vskip2mm
\noindent
Let us begin with describing some general features which we observed for 
the zero modes on the torus. We find
that the zero modes show a single lump in both densities $\rho$ and $\rho_5$.
For all values of $\zeta$ we find that the 
lumps are localized in space and time and on the average are rotationally
invariant, i.e.\ no space-time direction is singled 
out by the shape of the lumps. 

\begin{figure}[t]
\begin{center}
\epsfig{file=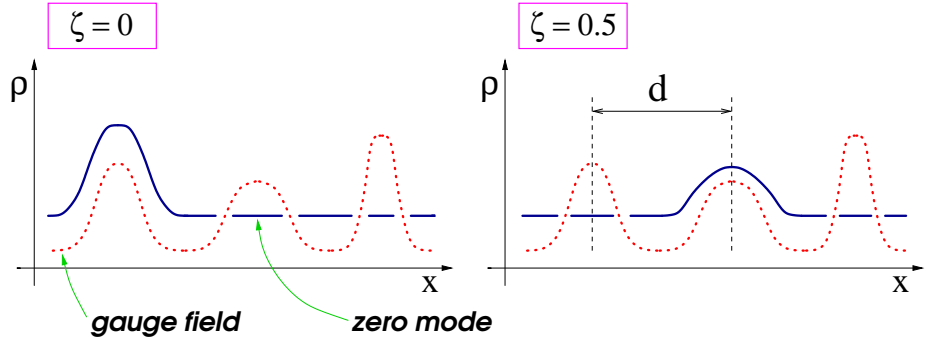,height=4.5cm,clip}
\end{center}
\vspace{-5mm}
\caption{{\sl Schematic sketch of the motion of the zero mode when switching 
the fermionic boundary condition from $\zeta = 0$ (l.h.s.) to 
$\zeta = 0.5$ (r.h.s.).}
\label{bumpsketch}}
\end{figure}

When we change the boundary condition the number of zero modes remains 
invariant, i.e.\ we always find exactly 1 zero 
mode\footnote{We remark that for both lattice sizes we found for 
about 1\% of the configurations an exception, i.e.\ a single value of 
$\zeta$ where there was no or two zero modes.}.  
When we now compare the zero modes with different boundary conditions but on 
the same configurations we often find drastic differences: The position
of the lump in space time can be different and also the inverse participation
ratio. In Fig.\ \ref{bumpsketch} we show a schematic plot of the typical 
behavior often observed when comparing the zero mode for periodic and 
anti-periodic boundary conditions. 

In the two plots of Fig.\ \ref{bumpsketch} on the horizontal axes we show 
all of space time - the corresponding variable is $x$. On the vertical axes 
we plot some general density $\rho$. The zero mode is represented by the 
full curve and in this case the density $\rho$ on the vertical axes is 
the scalar density defined in Eq.\ (\ref{scalardens}). The left hand side 
(l.h.s.) plot shows the case of periodic boundary conditions ($\zeta = 0.0$) 
while the r.h.s.\ plot is for anti-periodic boundary conditions 
($\zeta = 0.5$). For the different boundary conditions the zero mode
is located at different positions in space time. When scanning through the 
boundary parameter $\zeta$ in small steps of $0.1$ we found that the zero mode
can be located at up to three different positions. To be more specific, for 
a subensemble of 10 configurations on the $16^4$ lattice 
where we computed eigenvectors with small 
steps of $\zeta$ we found that 4 zero modes visited 3 different locations, 
2 zero modes visited 2 locations and 4 zero modes remained at the 
same position for all values of $\zeta$. 

For cooled configurations it has been established that lumps seen by the 
zero modes are associated with localized structures in the gauge 
fields. For thermalized configurations quantum fluctuations are present, but 
the zero modes and also the low lying modes
couple only to the infrared structures in the gauge 
field \cite{chiralitypapers}. 
In Fig.\ \ref{bumpsketch} we indicate such localized lumps in the infrared
part of the gauge field
using dashed curves. Here the density on the vertical axis is e.g.\ the 
gluonic action density a quantity that is often used in the study of 
gauge lumps in cooled configurations. In the r.h.s.\ plot we also show
the distance $d$ (compare Eq.\ (\ref{distance})) 
between the lump seen by the zero mode with
periodic boundary conditions and the anti-periodic lump.

We stress, that when the zero mode visits different 
lumps when changing $\zeta$,
the lumps always have the same chirality. When inspecting the 
pseudoscalar density $\rho_5$ we find that all lumps are 
seen with the same sign 
of $\rho_5$. This is in agreement with the index theorem, which implies
that if there is only one zero mode, its chirality is given by minus
the topological charge. Since we do not touch the gauge configuration
and only change $\zeta$, the index theorem requires the chirality 
of the zero mode to be the same for all fermionic boundary conditions. 

\begin{figure}[t]
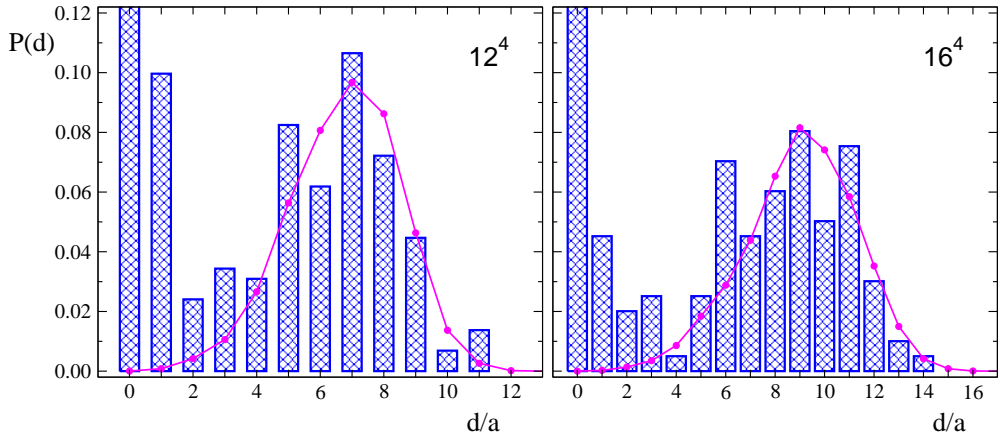

\begin{center}
\hspace*{-9mm}
\epsfig{file=disthistoz12b845.eps,height=5.7cm,clip}
\hspace{-2mm} 
\epsfig{file=disthistoz16b845.eps,height=5.7cm,clip}
\end{center} 
\vspace{-6mm}
\caption{{\sl
Distributions of the distance between the peaks in the zero modes
seen with 
periodic and anti-periodic boundary conditions. The distance is given in
lattice units. The l.h.s.\ plot is for lattice size $12^4$ (291 configurations), 
the r.h.s.\ plot for $16^4$ (199 configurations). 
The curve represents the numbers from a 
simple model to describe the data (see the text for details).} 
\label{disthisto}}
\end{figure}

Let us discuss an important consistency check we performed. The boundary 
conditions for the gauge fields are periodic in all directions. In order 
to use the 
zero modes as an analyzing tool we single out one of the directions by 
applying the boundary condition. However, the infrared structures of the 
gauge field should not depend on this choice. In order to check this, we 
applied the non-trivial fermionic boundary conditions not only in the 
4-direction, but compared also with the zero-modes computed with the 
boundary condition applied to one of the other directions. We find that the
same positions for the constituents are identified when applying the boundary
conditions in different directions. 
\vskip5mm
\noindent
{\bf Systematic comparison of periodic and anti-periodic zero modes}
\vskip2mm
\noindent
Let us now come to presenting the results of a large scale comparison 
of anti-periodic and periodic zero modes. We begin with a discussion of 
the distribution of the distance $d$ (compare Eq.\ (\ref{distance})) 
between the peaks in the periodic and anti-periodic modes. The corresponding 
histograms for our $12^4$ ensemble are shown on the l.h.s.\ 
and for the $16^4$ ensemble on the r.h.s.\ of Fig.\ \ref{disthisto}. 
The distance is measured in 
lattice units $a = 0.094(1)$ fm and the histograms are normalized to 1.

For both lattice sizes we find that the histogram has large contributions
for very small distances between $d = 0$ and $d = 2a$. These contributions 
come from configurations where the periodic and the anti-periodic zero mode
sit at essentially the same position, with the small discrepancies in the
position partly due to quantum fluctuations. Clearly separated from these modes
we find a well pronounced peak in the histogram 
containing configurations which show 
distances of at least $d = 3a$. For both volumes the number of 
configurations contributing to this peak is about half of 
the total statistics.  

An interesting question is whether one can understand the  
distribution of distances and the peak of the histograms. 
We experimented with a simple
model where the positions of the two lumps are distributed independently on 
the lattice. In $\mbox{I\hspace{-0.8mm}R}^4$ such a distribution would
be proportional to $d^3$ but on the toroidal lattice we obtain a distribution 
which has a maximum at $L/2 +1$ where $L$ is the linear extension of the 
lattice. The corresponding curves are shown in the plots as connected dots. 
Note that these model curves are not normalized to 1. Instead their 
normalization was chosen such that they optimally describe 
(using a $\chi^2$ optimization) the data in the range $d \geq 2a$. 
Using this procedure we find
that the area under the model curve is 0.425 for $12^4$ and 0.439 for $16^4$. 
For both volumes the model describes reasonably well 
the part of the histogram coming from separated periodic and anti-periodic
zero modes. This implies that at least 40 percent of the charge 
$\pm 1$ configurations show two or more locations for their
zero mode and the corresponding lumps seem to be distributed independently.
Note that we only compared periodic and anti-periodic boundary conditions, 
and we thus do not count cases where the periodic and anti-periodic
zero modes are located at the same position, but a different position
is visited for some $\zeta \neq 0.0, 0.5$. We thus expect that the fraction
of configurations with ``jumping'' zero modes is actually larger than 40 
percent.
\begin{figure}[t]
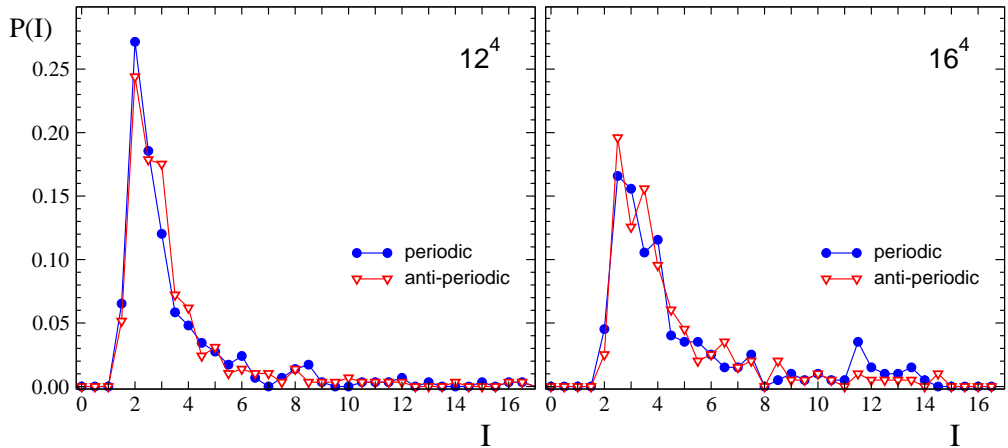

\begin{center}
\hspace*{-8mm}
\epsfig{file=iprhistoz12b845.eps,height=5.9cm,clip}
\hspace{-2mm} 
\epsfig{file=iprhistoz16b845.eps,height=5.9cm,clip}
\end{center} 
\vspace{-7mm}
\caption{{\sl Distributions of the inverse participation ratio for the
zero modes. We compare the distribution for periodic (filled circles)
and anti-periodic (open triangles) boundary conditions.} 
\label{iprhisto}}
\end{figure}

In Fig.\ \ref{iprhisto} we compare the distribution of the 
inverse participation
ratio of the periodic zero modes to the distribution for the anti-periodic 
zero modes. The l.h.s.\ plot is for lattice size $12^4$, the
r.h.s.\ plot for $16^4$. For both volumes we find that the distribution peaks
at a relatively low value, with 
the distribution shifted to larger values of $I$ for $16^4$. The remarkable
feature of the plots is the good agreement of the distribution for
the periodic zero modes with the distribution for the anti-periodic modes. 
This agreement demonstrates again that the size distribution of the lumps
is governed by the infrared structures of the underlying gauge field 
and is not distorted by the zero modes with different boundary conditions
which we use as analyzing tool. Such a distortion could manifest itself
by a different distribution of $I$ for periodic and anti-periodic zero modes,
which is clearly ruled out by our data.
\vskip5mm
\noindent
{\bf The nature of the topological lumps}
\vskip2mm
\noindent   
In the previous sections 
we have demonstrated that the zero mode can be located at different
positions if one changes the fermionic boundary conditions. We have shown
that for at least 40\% of the configurations the maximum of the scalar density
visits two or more different space-time points. 
The zero modes are localized in all 4 directions and 
the chirality of the zero mode remains unchanged 
in accordance with the index theorem. 
The crucial question is now whether the zero modes hop between gauge lumps of
fractional charge or simply move from one charge $\pm 1$ object to another. 

Let us first try to find a possible 
mechanism for the latter scenario, where the hopping 
comes about through some intricate interplay of
independent topological objects with charge $\pm 1$. For 
example the appearance of three lumps 
with positive chirality seen by the zero mode 
could come from a configuration with 
three anti instantons and two instantons. Such a configuration would have 
net charge $-1$ and naively one would expect 5 zero modes. However, 
quantum fluctuations could distort 4 of the zero modes, shift their
eigenvalues up or down the imaginary axis and turn them into near zero modes. 
The remaining
single zero mode could then couple only to the three anti instantons
and in this way show only positive chirality for all three lumps. 

This scenario cannot be entirely ruled out, but arguments
were given in \cite{instantonoverlap} that such behavior is 
unlikely to occur, at least so at finite temperature.
Also our comparison
of two different volumes makes this scenario unlikely. If one increases the 
volume, more topological objects with charge $\pm 1$ can fit into our 
box. This would imply that the scenario based on the interplay of charge $\pm 1$
objects could happen more often for the larger volume. 
Our data show however, that
the rate remains close to 40\% for both volumes although the larger lattice
has a space-time volume more than three times as big as the smaller lattice. 
If on the other hand the hopping of the zero-modes is a genuine
property of charge $\pm 1$ configurations, caused e.g.\ by fractionally charged 
constituents, then the percentage of charge $\pm 1$ 
configurations showing hops should be independent of the volume. 
This is what we observe.

At this point we would like to comment on related work by
Ilgenfritz, Martemyanov, M\"uller-Preussker and Veselov \cite{berlin}. 
In their article SU(2) gauge fields at varying temperature are subjected
to a standard cooling technique. The authors concentrate on (approximate)
solutions of the lattice field equations with topological charge $\pm 1$.
Whereas in the confinement phase near the critical temperature a
considerable fraction of configurations is seen to consist of separated
constituents as static BPS monopoles, at lower temperatures topological
lumps of integer topological charge, i.e.\ non-dissociated calorons show
up. In terms of the Polyakov loop variable it turns out
that these configurations have an intrinsic ``$\pm 1$ dipole'' structure
connected to the nontrivial asymptotics of the holonomy.
The form of the solutions is in one-to-one correspondence
to the analytically known KvB solutions. In an analogous
manner they have also studied the case of the 4-torus. They have found
always non-dissociated ``instantons'' but with a similar dipole structure in
all space-time directions for which the asymptotic holonomy is
non-trivial. In this sense an interpretation of the topological lumps in
terms of constituents is proven also in their paper. 
The fact that these lumps are not seen dissolved into constituents
at $T=0$ is not straightforward to
reconcile with our constituent interpretation of the 
zero-mode hopping for equilibrium fields. The lack of separate
constituents in
the cooled configurations may be an artifact of the standard
cooling method which could drive the constituents into such a ``bound state''. 
Together with the authors of \cite{berlin} we are currently beginning
a detailed analysis of this issue 
using APE blocking and improved cooling techniques 
combined with fermionic methods. 
\vskip5mm
\noindent
{\bf Summary and outlook}
\vskip2mm
\noindent
In this article we show that for SU(3) gauge theory on the 
torus the zero mode of charge $\pm 1$ configurations 
can couple to different topological objects. 
In particular we find that the zero mode often is localized 
at different positions in space-time when changing the boundary conditions,
while the chirality of the lumps seen by the zero modes remains the same. 
For two different volumes we compared the periodic and the anti-periodic
zero modes for a large number of charge $\pm 1$ configurations. For both
volumes we find that for about 40 \% of the configurations the periodic and 
the anti-periodic zero modes are located at different positions. 
We compare the distribution of the distance between the
different lumps to a simple model and find that the behavior of the 
constituents is compatible with independent placement on the torus. 
 
Although we cannot entirely rule out the possibility 
that the hopping is between integer charged objects, we lean towards the
interpretation that it is true fractionally charged constituents we 
observe. This interpretation is supportet by analytic 
arguments at finite $T$ showing that hopping between integer lumps is 
unlikely to occur. Also the absence
of a volume dependence favors an explanation by fractionally charged 
constituents.

Our results support a possible extension of the monopole constituent
picture known for manifolds with one or two compact dimensions to the 
4-torus. Such a generalization of the known classical solutions to the
torus (with twisted boundary conditions)
is certainly a challenging and interesting problem. On the numerical 
side also several interesting questions remain to be studied. In order to 
avoid problems with mixing of zero modes we have so far restricted ourselfes 
to configurations with exactly one zero mode. It would be interesting 
to analyze also the zero modes for higher topological sectors and check 
whether a similar behavior can be identified. Another direction for
research would be an analysis of the near zero modes. They are expected 
to emerge when topological objects start to overlap. For the lowest 
modes one expects that the behavior of the lumps seen by the eigenvectors
still resembles the properties of the exact zero modes. 
Showing constituent-like
behavior for these near zero modes, which through the Banks-Casher relation 
are responsible for chiral symmetry breaking, would lead one step closer to
a possible unification of the mechanisms for confinement and chiral 
symmetry breaking.

\newpage
\noindent
{\bf Acknowldegements:} 
We thank Falk Bruckmann, 
Meinulf G\"ockeler, Mi\-cha\-el Ilgenfritz, 
Christian Lang, Kurt Langfeld, Dirk Peschka, 
Michael M\"ul\-ler-Preussker, Hugo Reinhardt, 
Stefan Schaefer, Andreas Sch\"afer, Stefan Solbrig and Pierre van Baal
for discussions. The calculations were done on the Hitachi SR8000
at the Leibniz Rechenzentrum in Munich and we thank the LRZ staff for
training and support. This work was supported by the DFG Forschergruppe
``Lattice-Hadron-Phenomenology''.


\begin{thebibliography}{1234567}

\bibitem{kvb}
T.C.~Kraan and P.~van~Baal, 
Phys.\ Lett.\ B 428 (1998) 268,
Phys.\ Lett.\ B 435 (1998) 389,
Nucl.\ Phys.\ B 533 (1998) 627;
K. Lee and C. Lu,
Phys.\ Rev.\ D 58 (1998) 1025011.

\bibitem{highercharge}
F.~Bruckmann and P.~van~Baal, 
Nucl.\ Phys.\ B 645 (2002) 105;
F.\ Bruckmann, D.\ Nogradi and P.\ van Baal,
Acta Phys.\ Polon.\ B 34 (2003) 5717
(hep-th/0309008).

\bibitem{kvblattice1}
M.\ Garcia P\'erez, A.\ Gonzalez-Arroyo, A.\ Montero and P.\ van Baal,
JHEP 9906 (1999) 001.

\bibitem{kvblattice2}
E.-M. Ilgenfritz, B.~V. Martemyanov, M. M\"uller-Preussker, 
S.~Shcheredin and A.~I.~Veselov, 
Phys.\ Rev.\ D 66 (2002) 074503,
Nucl.\ Phys.\ Proc.\ Suppl.\ 119 (2003) 754.

\bibitem{kvblattice3}
C.~Gattringer, 
Phys.\ Rev.\ D 67 (2003) 034507.

\bibitem{kvblattice4}
C.~Gattringer and S.~Schaefer, 
Nucl.\ Phys.\ B 654 (2003) 30.

\bibitem{kvbzeromodes} 
M.~Garcia P\'erez, A.~Gonz\'alez-Arroyo, C.~Pena and P.~van Baal, 
Phys.\ Rev.\ D 60 (1999) 031901;
M.N.~Chernodub, T.C.~Kraan and P.~van Baal, 
Nucl.\ Phys.\ Proc.\ Suppl.\ 83 (2000) 556;
F.\ Bruckmann, D.\ Nogradi, P.\ van Baal,
Nucl.\ Phys.\ B 666 (2003) 197.

\bibitem{selfdual}
C.~Gattringer,
Phys.\ Rev.\ Lett.\ 88 (2002) 221601.

\bibitem{kvblattice5}
C.\ Gattringer, E.-M.\ Ilgenfritz, B.V.\ Martemyanov, M.\ 
M\"uller-Preu{\ss}ker, 
D.\ Peschka, R.\ Pullirsch, S.\ Schaefer and A.\ Schafer,
hep-lat/0309106.

\bibitem{t4analytic}
A.~Gonzalez-Arroyo,
{\sl Yang-Mills fields on the 4-dimensional torus. (Classical theory)}, in:
{\sl Advanced school of non-perturbative quantum field physics},
M.\ Asorey and A.\ Dobado (eds), World Scientific, 1998
(hep-th/9807108);
P.\ van Baal, {\sl QCD in a finite volume},
in: {\sl At the Frontiers of Particle
Physics -- Handbook of QCD, Boris Ioffe Festschrift}, Vol.2,
M.\ Shifman (ed), World Scientific, 2001 
(hep-ph/0008206).

\bibitem{fopa}
C.\ Ford, J.M.\ Pawlowski, T.\ Tok and A.\ Wipf,
Nucl.\ Phys.\ B 596 (2001) 387;
C.~Ford and J.~M.~Pawlowski,
Phys.\ Lett.\ B 540 (2002) 153 and 
hep-th/0302117.

\bibitem{luweact}
M.~L{\"u}scher and P.~Weisz, Commun.~Math.~Phys.~97 (1985) 59; 
Err.: 98 (1985) 433; 
G.~Curci, P.~Menotti and G.~Paffuti, Phys.~Lett.~B 130 (1983) 205, 
Err.: B 135 (1984) 516.

\bibitem{scale}
C.~Gattringer, R.~Hoffmann and S.~Schaefer,
Phys.\ Rev.\ D 65 (2002) 094503.

\bibitem{chirimp} 
C.~Gattringer, Phys.~Rev.~D 63 (2001) 114501;
C.~Gattringer, I.~Hip, C.B.~Lang, Nucl.~Phys.~B597 (2001) 451.

\bibitem{instantontest}
C.\ Gattringer, M.\ G\"ockeler, C.B.\ Lang, P.E.L.\ Rakow and A.\ Sch\"afer,
Phys.\ Lett.\ B 522 (2001) 194.

\bibitem{arnoldi}
D.C.\ Sorensen, SIAM J.Matrix Anal.Appl.13 (1992) 357.

\bibitem{chiralitypapers}
T.L.\ Ivanenko and J.W.\ Negele,
Nucl.\ Phys.\ Proc.\ Suppl.\ 63 (1998) 504;
%
T.\ DeGrand and A.\ Hasenfratz,
Phys.\ Rev.\ D 64 (2001) 034512;
%
T.\ DeGrand,
Phys.\ Rev.\ D 64 (2001) 094508;
%
I.\ Hor\-v\'ath {\it et al.},
Phys.\ Rev.\ D 67 (2003) 011501,
Phys.\ Rev.\ D 65 (2002) 014502;
%
T.\ DeGrand and A.\ Hasenfratz,
Phys.\ Rev.\ D 65 (2002) 014503;
%
T.\ Blum {\it et al.},
Phys.\ Rev.\ D 65 (2002) 014504;
%
R.G.\ Edwards and U.M.\ Heller,
Phys.\ Rev.\ D 65 (2002) 014505;
%
I.\ Hip {\it et al.},
Phys.\ Rev.\ D 65 (2002) 014506;
%
C.\ Gattringer {\it et al.},
Nucl.\ Phys.\ B 617 (2001) 101, (1985)
Nucl.\ Phys.\ B 618 (2001) 205,
Nucl.\ Phys.\ Proc.\ Suppl.\  106 (2002) 492,
Nucl.\ Phys.\ Proc.\ Suppl.\  106 (2002) 551;
%
P.\ Hasenfratz {\it et al.}, 
Nucl.\ Phys.\ Proc.\ Suppl.\  106 (2002) 751.

\bibitem{instantonoverlap}
F.~Bruckmann, M.~Garcia P\'erez, D.~Nogradi and P.~van~Baal, hep-lat/0308017.

\bibitem{berlin}
E.-M.\ Ilgenfritz, B.V.\ Martemyanov, M.\ M\"uller-Preussker
and A.I.\ Veselov, {\it Recombination of dyons into calorons
in SU(2) lattice fields at low temperature},
to appear in eprint-Archive hep-lat.

\end{thebibliography}
\end{document}